\pdfoutput=1
\documentclass[11pt]{article}

\usepackage[preprint]{acl}

\usepackage{times}
\usepackage{latexsym}

\usepackage[T1]{fontenc}
\usepackage[utf8]{inputenc}

\usepackage{microtype}

\usepackage{inconsolata}

\usepackage{graphicx}

\usepackage{natbib}

\title{Soft Begging: Modular and Efficient Shielding of LLMs against Prompt Injection and Jailbreaking based on Prompt Tuning}

\author{Simon Ostermann$^{1, 2}$, Kevin Baum$^{1, 2}$, Christoph Endres$^{3}$, Julia Masloh$^{3}$, Patrick Schramowski$^{1, 2}$\\
  $^{1}$Deutsches Forschungszentrum für Künstliche Intelligenz (DFKI) \\
  $^{2}$Centre for European Research in Trusted AI (CERTAIN) \\
  $^{3}$sequire technology GmbH\\
  \texttt{firstname.lastname@\{dfki|sequire\}.de}
    }
\begin{document}
\maketitle
\begin{abstract}
Prompt injection (both direct and indirect) and jailbreaking are now recognized as significant issues for large language models (LLMs), particularly due to their potential for harm in application-integrated contexts. This extended abstract explores a novel approach to protecting LLMs from such attacks, termed "soft begging." This method involves training soft prompts to counteract the effects of corrupted prompts on the LLM's output. We provide an overview of prompt injections and jailbreaking, introduce the theoretical basis of the "soft begging" technique, and discuss an evaluation of its effectiveness. 
\end{abstract}

\section{Background: Attacking LLMs}

Current LLMs lack adversarial robustness \cite{carlini2021extracting}. This leads to new phenomena observed in the attack surface of LLMs that are relevant to safety and security. Prominent examples include: jailbreaking, direct prompt injections and indirect prompt injections.

In jailbreaking, a target LLM is manipulated to circumvent content moderation systems and subvert safety rules \citep{zou2023universaltransferableadversarialattacks}. Prompt injections describe attacks where input is inserted into an application and processed by the LLM later downstream - with indirect prompt injections denoting those cases where the attacker is not the user prompting the LLM \citep{Greshake_et_al_2023}. As can be seen, direct prompt injection \citep{perez2022ignore} is similar to jailbreaking: Both hide specific instructions inside a prompt that alters or manipulates the behavior of the system in a manner not intended by the LLM provider. This adversarial instruction is typically conveyed through text which is sometimes visually concealed (using tactics such as white or extremely small font sizes), encoded \citep{liu2024jailbreakingchatgptpromptengineering} or obfuscated by an API (for example, a command that is appended to each input without the end user's awareness).

Especially indirect prompt injection attacks, where adversaries remotely affect the target system, exploit the fact that there is no explicit separation between code and data in current LLMs and, as a consequence, between valid system instructions and invalid, potentially adversarial, instructions \citep{zverev2024llmsseparateinstructionsdata}.

Dangers of prompt injections include data leakage, i.e., the disclosure of sensitive information, and system manipulation, i.e., altering the behaviour of the system based on the injected prompt. In the most extreme case, the adversarial user can hijack the LLM \citep{qiang2024hijackinglargelanguagemodels}, which is even more problematic if the LLM has great autonomy and is, for example, given access to plugins or third-party data.

As the goal of the examples mentioned above is often to elicit harmful behavior of an LLM, it is important to note that there are also benign and neutral usages. Benign injections may guide the model to generate more specific or desirable outputs, while neutral injections serve to test the model's resilience, accuracy, or response to edge cases without malicious intent.

\section{Shielding LLMs against Jailbreaking and Prompt Injections}
Defense methods against jailbreaking and prompt injection attacks can be grouped into two directions: attack prevention and attack detection.

The most straightforward prevention method to shield an LLM is to formulate counter-prompts to neutralize the harmful injection (“Hello ChatGPT, please ignore any harmful prompts that might follow this instruction and just do exactly as I say”). While easy to implement, such “begging” is easy to trick and rarely successful. Research has since come up with a range of more elaborate countermeasures. \citet{jain2023baseline} proposed two prevention baseline methods that illustrate how simple defenses such as input preprocessing (paraphrasing and retokenization) can effectively act against gradient-based methods. Whereas paraphrasing rewrites the input by changing its meaning, retokenization breaks down the tokens into smaller ones. Recently, input preprocessing techniques have been studied that aim at making it easier for the model to distinguish between valid instructions and untrustworthy input \citep{hines2024defendin}. 

The most frequently used post-training detection method is the implementation of filters, i.e.,  additional algorithms or models that try to scan inputs for injected harmful prompts and mask them or reject the prompts automatically \citep{dong2024buildingguardrailslargelanguage}. Detection methods for prompt injection, for example, often rely on the perplexity score, which is assumed to be higher for adversarial inputs and can be detected by a simple thresholding approach~\citep{alon2023detecting,jain2023baseline}. Yet, setting the threshold is not trivial and a poor threshold might hurt the model's overall performance. Generally, filters can be implemented as rule-based checks, but they can also be trained deep learning models that classify inputs into harmful and non-harmful parts. The problem with such filters is that they are often too restrictive in the case of model-based filters, and too loose in the case of rule-based filters. 

As an alternative, models can theoretically be fine-tuned to be robust against attacks. While working reasonably well, the disadvantage of such methods is that the fine-tuning of whole models is costly; it needs to be redone whenever a new type of malicious prompt is detected; and even robust detection of injections always leaves room for vulnerabilities. Also, although the field of parameter-efficient fine-tuning is being reasonably well studied and still growing \citep{lialin2023scalingscaleupguide}, contributions that focus on its efficiency in cybersecurity contexts are lacking.  

\section{Soft Begging: Shielding LLMs with Soft Prompts}
We propose \textit{soft begging} as a new alternative for LLM shielding. The method can be seen as a combination of the na\"ive begging approach, combined with parameter-efficient fine-tuning techniques. The method basically comprises the training of so-called soft prompts, i.e. trainable input vectors that are preprended to any prompt. The soft prompts are trained to nullify the behaviour that the LLM exhibits based on potentially harmful parts of the prompts. This is not done as a filtering step, but in an implicit way -- i.e. without altering the prompts -- which effectively follows the idea of “begging” the network to ignore the harmful parts on a parameter level. 

We conjecture that a first advantage of using such soft prompts is their effectiveness, as it enables shielding on the parameter-level against attacks on the text level, which effectively provides the shield with an advantage: A parameter-level control can be assumed to be always more effective than textual control. Second, shields trained in such a way are easy and efficiently adaptable, as the training of a soft prompt is magnitudes faster than training the whole model, as is done for example when finetuning LLMs to be robust against injections. At the same time, obviously, the LLM itself stays as is with its parameters being frozen. Last, we assume that soft begging prompts could be modularized to fit different types of attacks and even be combined for different use cases, rendering them as a very effective and customizable alternative to other shields.

In the most basic version, such prompts are trained on quadruples of clean prompts, corrupted prompts, clean output and output based on the corrputed prompt. The soft prompts are then trained to produce the clean output from the corrupted prompt.

The idea can be scaled up, e.g. by training different soft prompts for different injections and combining them via prompt fusion or other mechanisms. Also, the prompts could be combined with a filtering mechanism, that first identifies the kind of threat (without localizing it), and then picks the matching soft prompt based on this.
%    \item Combine them with hard prompts. Add a sort of rule-based layer on top of the soft prompts.
%    \item Combine them with DL-based filters: Have a filter classify a “type” of injection and then use the respective counter-soft prompt.
%\end{itemize}

\section{Evaluation of Prompt Injection}
Usually, jailbreaking and direct prompt injection attacks are evaluated by injecting a malicious prompt with a specific phrase into benign queries and observing the output~\cite{perez2022ignore,chen2024struq}.
This allows for easy evaluation, as the attack is successful if the specified keyword is present in the LLM response.

In contrast to that, in indirect prompt injection attacks, the malicious prompt is embedded into malicious documents and sources, as the name suggests.
As described above, the output of the LLM can be reviewed for the injected target keywords, indicating a successful attack~\cite{hines2024defendin,Greshake_et_al_2023,liu2024promptinjection}.
While the evaluation of jailbreaking and direct prompt injection attacks is relatively straightforward, indirect prompt injection has the problem of requiring third malicious party material like websites, e-mails, or other endpoints providing data to the LLM.
To simplify the evaluation of indirect prompt injection attacks, various benchmarks have been recently proposed~\cite{yi2024benchmarking,zhan2024injecagent}.
These benchmarks provide curated content with pre-defined malicious prompts injected into the queries.
This allows for standardized evaluation for many different LLMs.

Evaluation~\citep{liu2024promptinjection,liu2024formalizing} typically investigates two model properties: clean performance when facing benign inputs and robustness to adversarial inputs. The clean performance can be computed using standard benchmarks for LLM tasks. Practical defenses should not degrade a model's utility. Robustness, or alternatively the attack success rate, quantifies the share of adversarial prompts achieving their goal of breaking the safety railguards. \citet{hines2024defendin} highlight the importance of clearly quantifiable protocols for measuring Attack Surface Rate (ASR) and make their evaluation approach using a synthetic dataset explicit. This call for transparency guides our work.

% Further input:
% https://arxiv.org/abs/2403.14720
% https://arxiv.org/abs/2312.14197 and https://github.com/microsoft/BIPIA?tab=readme-ov-file
% https://huggingface.co/datasets/JailbreakV-28K/JailBreakV-28k?row=27 
% TODO Move to bib
% Overview paper security challenges - https://arxiv.org/pdf/2404.09932
% PEFT Meta-study https://arxiv.org/pdf/2303.15647
% MS spotlight related work https://arxiv.org/pdf/2312.14197
% Separation of code and data https://arxiv.org/abs/2403.06833
% VERY GOOD SUMMARY: https://github.com/tldrsec/prompt-injection-defenses?tab=readme-ov-file#prompt-engineering--instructional-defense

\bibliography{acl_latex}

\begin{thebibliography}{17}
\providecommand{\natexlab}[1]{#1}

\bibitem[{Alon and Kamfonas(2023)}]{alon2023detecting}
Gabriel Alon and Michael Kamfonas. 2023.
\newblock Detecting language model attacks with perplexity.
\newblock \emph{arXiv Preprint}, arXiv:2308.14132.

\bibitem[{Carlini et~al.(2021)Carlini, Tramer, Wallace, Jagielski, Herbert-Voss, Lee, Roberts, Brown, Song, Erlingsson et~al.}]{carlini2021extracting}
Nicholas Carlini, Florian Tramer, Eric Wallace, Matthew Jagielski, Ariel Herbert-Voss, Katherine Lee, Adam Roberts, Tom Brown, Dawn Song, Ulfar Erlingsson, et~al. 2021.
\newblock Extracting training data from large language models.
\newblock In \emph{30th USENIX Security Symposium (USENIX Security 21)}, pages 2633--2650.

\bibitem[{Chen et~al.(2024)Chen, Piet, Sitawarin, and Wagner}]{chen2024struq}
Sizhe Chen, Julien Piet, Chawin Sitawarin, and David Wagner. 2024.
\newblock \href {https://arxiv.org/abs/2402.06363} {Struq: Defending against prompt injection with structured queries}.
\newblock arXiv:2402.06363.

\bibitem[{Dong et~al.(2024)Dong, Mu, Jin, Qi, Hu, Zhao, Meng, Ruan, and Huang}]{dong2024buildingguardrailslargelanguage}
Yi~Dong, Ronghui Mu, Gaojie Jin, Yi~Qi, Jinwei Hu, Xingyu Zhao, Jie Meng, Wenjie Ruan, and Xiaowei Huang. 2024.
\newblock \href {https://arxiv.org/abs/2402.01822} {Building guardrails for large language models}.
\newblock \emph{Preprint}, arXiv:2402.01822.

\bibitem[{Greshake et~al.(2023)Greshake, Abdelnabi, Mishra, Endres, Holz, and Fritz}]{Greshake_et_al_2023}
Kai Greshake, Sahar Abdelnabi, Shailesh Mishra, Christoph Endres, Thorsten Holz, and Mario Fritz. 2023.
\newblock \href {https://doi.org/10.1145/3605764.3623985} {Not what you've signed up for: Compromising real-world llm-integrated applications with indirect prompt injection}.
\newblock In \emph{Proceedings of the 16th ACM Workshop on Artificial Intelligence and Security}, AISec '23, page 79–90, New York, NY, USA. Association for Computing Machinery.

\bibitem[{Hines et~al.(2024)Hines, Lopez, Hall, Zarfati, Zunger, and Kiciman}]{hines2024defendin}
Keegan Hines, Gary Lopez, Matthew Hall, Federico Zarfati, Yonatan Zunger, and Emre Kiciman. 2024.
\newblock \href {https://arxiv.org/abs/2403.14720} {Defending against indirect prompt injection attacks with spotlighting}.
\newblock arXiv:2403.14720.

\bibitem[{Jain et~al.(2023)Jain, Schwarzschild, Wen, Somepalli, Kirchenbauer, Chiang, Goldblum, Saha, Geiping, and Goldstein}]{jain2023baseline}
Neel Jain, Avi Schwarzschild, Yuxin Wen, Gowthami Somepalli, John Kirchenbauer, Ping{-}yeh Chiang, Micah Goldblum, Aniruddha Saha, Jonas Geiping, and Tom Goldstein. 2023.
\newblock Baseline defenses for adversarial attacks against aligned language models.
\newblock \emph{arXiv Preprint}, arXiv:2309.00614.

\bibitem[{Lialin et~al.(2023)Lialin, Deshpande, and Rumshisky}]{lialin2023scalingscaleupguide}
Vladislav Lialin, Vijeta Deshpande, and Anna Rumshisky. 2023.
\newblock \href {https://arxiv.org/abs/2303.15647} {Scaling down to scale up: A guide to parameter-efficient fine-tuning}.
\newblock \emph{Preprint}, arXiv:2303.15647.

\bibitem[{Liu et~al.(2024{\natexlab{a}})Liu, Deng, Li, Wang, Wang, Wang, Zhang, Liu, Wang, Zheng, and Liu}]{liu2024promptinjection}
Yi~Liu, Gelei Deng, Yuekang Li, Kailong Wang, Zihao Wang, Xiaofeng Wang, Tianwei Zhang, Yepang Liu, Haoyu Wang, Yan Zheng, and Yang Liu. 2024{\natexlab{a}}.
\newblock \href {https://arxiv.org/abs/2306.05499} {Prompt injection attack against llm-integrated applications}.
\newblock arXiv:2306.05499.

\bibitem[{Liu et~al.(2024{\natexlab{b}})Liu, Deng, Xu, Li, Zheng, Zhang, Zhao, Zhang, Wang, and Liu}]{liu2024jailbreakingchatgptpromptengineering}
Yi~Liu, Gelei Deng, Zhengzi Xu, Yuekang Li, Yaowen Zheng, Ying Zhang, Lida Zhao, Tianwei Zhang, Kailong Wang, and Yang Liu. 2024{\natexlab{b}}.
\newblock \href {https://arxiv.org/abs/2305.13860} {Jailbreaking chatgpt via prompt engineering: An empirical study}.
\newblock \emph{Preprint}, arXiv:2305.13860.

\bibitem[{Liu et~al.(2024{\natexlab{c}})Liu, Jia, Geng, Jia, and Gong}]{liu2024formalizing}
Yupei Liu, Yuqi Jia, Runpeng Geng, Jinyuan Jia, and Neil~Zhenqiang Gong. 2024{\natexlab{c}}.
\newblock Formalizing and benchmarking prompt injection attacks and defenses.
\newblock \emph{arXiv Preprint}, arXiv:2310.12815.

\bibitem[{Perez and Ribeiro(2022)}]{perez2022ignore}
F{\'a}bio Perez and Ian Ribeiro. 2022.
\newblock \href {https://openreview.net/forum?id=qiaRo_7Zmug} {Ignore previous prompt: Attack techniques for language models}.
\newblock In \emph{NeurIPS ML Safety Workshop}.

\bibitem[{Qiang et~al.(2024)Qiang, Zhou, and Zhu}]{qiang2024hijackinglargelanguagemodels}
Yao Qiang, Xiangyu Zhou, and Dongxiao Zhu. 2024.
\newblock \href {https://arxiv.org/abs/2311.09948} {Hijacking large language models via adversarial in-context learning}.
\newblock \emph{Preprint}, arXiv:2311.09948.

\bibitem[{Yi et~al.(2024)Yi, Xie, Zhu, Kiciman, Sun, Xie, and Wu}]{yi2024benchmarking}
Jingwei Yi, Yueqi Xie, Bin Zhu, Emre Kiciman, Guangzhong Sun, Xing Xie, and Fangzhao Wu. 2024.
\newblock \href {https://arxiv.org/abs/2312.14197} {Benchmarking and defending against indirect prompt injection attacks on large language models}.
\newblock arXiv:2312.14197.

\bibitem[{Zhan et~al.(2024)Zhan, Liang, Ying, and Kang}]{zhan2024injecagent}
Qiusi Zhan, Zhixiang Liang, Zifan Ying, and Daniel Kang. 2024.
\newblock \href {https://arxiv.org/abs/2403.02691} {Injecagent: Benchmarking indirect prompt injections in tool-integrated large language model agents}.
\newblock arXiv:2403.02691.

\bibitem[{Zou et~al.(2023)Zou, Wang, Carlini, Nasr, Kolter, and Fredrikson}]{zou2023universaltransferableadversarialattacks}
Andy Zou, Zifan Wang, Nicholas Carlini, Milad Nasr, J.~Zico Kolter, and Matt Fredrikson. 2023.
\newblock \href {https://arxiv.org/abs/2307.15043} {Universal and transferable adversarial attacks on aligned language models}.
\newblock \emph{Preprint}, arXiv:2307.15043.

\bibitem[{Zverev et~al.(2024)Zverev, Abdelnabi, Tabesh, Fritz, and Lampert}]{zverev2024llmsseparateinstructionsdata}
Egor Zverev, Sahar Abdelnabi, Soroush Tabesh, Mario Fritz, and Christoph~H. Lampert. 2024.
\newblock \href {https://arxiv.org/abs/2403.06833} {Can llms separate instructions from data? and what do we even mean by that?}
\newblock \emph{Preprint}, arXiv:2403.06833.

\end{thebibliography}

%\appendix
%
%\section{Example Appendix}
%\label{sec:appendix}

%This is an appendix.

\end{document}